\documentclass[conference]{IEEEtran}
\IEEEoverridecommandlockouts
\usepackage{cite}
\usepackage{amsmath,amssymb,amsfonts}
\usepackage{algorithmic}
\usepackage{graphicx}
\usepackage{textcomp}
\usepackage{xcolor}
\definecolor{light-gray}{gray}{0.95}
\usepackage{url}
\def\BibTeX{{\rm B\kern-.05em{\sc i\kern-.025em b}\kern-.08em
    T\kern-.1667em\lower.7ex\hbox{E}\kern-.125emX}}

\usepackage{minted}
\usepackage{flushend}
\usepackage{cite}
\usepackage{amsmath,amssymb,amsfonts}
\usepackage{algorithmic}
\usepackage{graphicx}
\usepackage{textcomp}
\usepackage{hyperref}
\usepackage{color}
\usepackage{booktabs}
\usepackage{xspace}
\usepackage{graphicx}
\usepackage{multirow}
\usepackage[framemethod=TikZ]{mdframed}
\usepackage{textcomp}
\usepackage{booktabs}
\usepackage{colortbl}
\usepackage{ragged2e}
\usepackage{courier}
\usepackage{textcomp} 
\usepackage{listings}
\usepackage{varwidth}
\usepackage{subcaption}

\def\BibTeX{{\rm B\kern-.05em{\sc i\kern-.025em b}\kern-.08em
    T\kern-.1667em\lower.7ex\hbox{E}\kern-.125emX}}

\usepackage[normalem]{ulem}

\newboolean{showcomments}
\setboolean{showcomments}{true}
\ifthenelse{\boolean{showcomments}}
	{\newcommand{\nb}[3]{
		{\colorbox{#2}{\bfseries\sffamily\scriptsize\textcolor{white}{#1}}}
		{\textcolor{#2}{\sf\small$\blacktriangleright$\textit{#3}$\blacktriangleleft$}}}
	 }
	{\newcommand{\nb}[3]{}
	 }

\usepackage{xcolor}
\usepackage{textcomp}
\usepackage[T1]{fontenc}
\usepackage{inconsolata}

\newcounter{rmd}

\begin{document}

\title{Towards Actionable Visualization: Ten Years Later, What Generative AI Changes and What It Cannot
}

\author{\IEEEauthorblockN{Leonel Merino}
\IEEEauthorblockA{School of Design, 
School of Engineering\\
Pontificia Universidad Católica de Chile\\
Santiago, Chile}
\and
\IEEEauthorblockN{Mohammad Ghafari}
\IEEEauthorblockA{Tehran Institute for Advanced Studies (TEIAS)\\ 
Khatam University\\
Tehran, Iran }
\and
\IEEEauthorblockN{Oscar Nierstrasz}
\IEEEauthorblockA{Feenk\\\\
Bern, Switzerland}
}

\maketitle

\begin{abstract}
Ten years ago, we surveyed software visualization research under the assumption that the challenge for adoption was matching developers' needs with techniques. Generative AI might have made that assumption obsolete by driving the cost of producing a visualization on demand toward zero. We argue that this mirrors a broader inversion already underway in software engineering: as AI devalues the production of artifacts, it elevates the human work of perceiving and directing them.
Looking back at our 2016 research, we found that domains we flagged as neglected, such as rationale, are exactly the ones AI now makes tractable, and a tool-sustainability problem we diagnosed then is one generative AI may worsen before it helps solve. One finding shifted outright: the share of studies delivered through immersive environments grew roughly fifteen-fold, though it remains a small minority of the field.
We argue that future research should focus on helping people understand large and complex software systems, including the reasoning processes generated by AI agents alongside the code itself.
Humans have always been accountable for what they trust, but AI may lead them to trust without enough scrutiny.
We confront the risks that follow, from the reliability of AI-generated visualization itself to the temptation of competing with machines on their own terrain, and 
conclude that actionable visualization is no longer a distant destination but a standard the field can now meet.
\end{abstract}

\begin{IEEEkeywords}
Software Visualization, Generative AI, Human Perception, Explainable AI, Program Comprehension
\end{IEEEkeywords}

\section{Introduction}
\label{sec:introduction}
Ten years ago, we surveyed 65 design studies from SOFTVIS/VISSOFT and asked two questions: which developer needs does visualization support, and how well~\cite{merino2016towards}. That paper received the VISSOFT 2016 best paper award, and an invited journal extension~\cite{merino2018towards} broadened the survey to 86 design studies. Both assumed the bottleneck was matching: developers had questions, the field had techniques, and adoption failed because the two rarely met. We called that survey \emph{Towards} Actionable Visualization, and the word \emph{towards} was deliberate; the field knew what a useful visualization looked like, but building one for a specific, momentary question cost too much to justify outside a sustained research effort. MetaVis, our own follow-up~\cite{merino2016metavis}, was built on the same assumption of scarcity: if developers could not find the right technique, we reasoned that a better catalog would help them match their needs to one.
That assumption no longer holds. Generative AI has not solved the matching problem the way we once tried to; it has collapsed the cost of producing a visualization in the first place, making the catalog largely beside the point. A developer with a question about a specific codebase can now have a disposable, purpose-built visualization generated on demand, rather than searching a taxonomy for the closest existing technique.

We read this through an analogy that is already reshaping how the field thinks about programming: AI has devalued the mechanical act of writing code while raising the value of reading, judging, and directing it~\cite{mckinsey2023economic}. The same inversion is underway in visualization, and we state it plainly:

\begin{quote}
\emph{Generating a visualization is becoming cheap. Seeing through one and judging what it shows is not, and will not be.}
\end{quote}

What survives the shift is not a nostalgic attachment to human effort. It is a specific, mechanical fact about vision: a person glancing at a bar chart sees which bar is tallest in a fraction of a second, before reading a single number, because the eye processes size and position in parallel. Reading the same comparison in a sentence takes longer and demands more attention, no matter how clearly the sentence is written~\cite{card1999using}. AI does not compete with that channel; it makes producing a visualization nearly free, handing the channel more to look at, including the complexity of AI's own making that was previously out of reach.


This paper is a self-critical revisit by the authors of the survey. Domains we flagged as neglected in 2016, rationale, contracts, and policies, turn out to be exactly the ones AI now makes tractable; the tool-lifespan problem we diagnosed then may be one that generative AI worsens before it helps solve; and the delivery medium, once almost entirely the standard screen, now includes a small but measurably growing share of immersive environments. Finally, we argue for a research agenda centered on human perceptual and evaluative capacity rather than technique production, and we conclude 
by examining the strongest objections that agenda has to survive.

\section{Revisiting the 2016 Landscape, a Decade Later}
\label{sec:revisiting}
Earlier, we classified 86 design studies along seven dimensions: task, need, audience, data source, representation, medium, and tool. 
Rereading that classification a decade later, three findings stand out: two are cases where our original data on under-served problem domains and on tool lifespan now tell a different story in light of generative AI, and one is a case where the data itself on delivery medium actually changed.

\subsection{Neglected domains, and the barrier AI removes: rationale, contracts, policies}
In the extended survey, we observed a mismatch between what developers ask and what visualization supports. 
We do not claim AI already resolves questions in these domains; we argue it removes the specific barrier, the cost of preparing heterogeneous data, that made them impractical to study at scale.
Domains such as \emph{rationale}, \emph{contracts}, and \emph{policies} were among the least visualized, yet ranked highly in developer need; domains such as \emph{architecture}, \emph{concurrency}, and \emph{dependencies} received comparatively more visualization attention than their reported importance warranted.
At the time, we read this as a gap to be addressed with more targeted design studies. In hindsight, it reads as an early indication of a barrier only AI would later remove: a question our earlier work classified under rationale, who wrote a given part of the system, and why, illustrates exactly that. Answering it required manually correlating version control metadata, commit messages, and source code structure, an effort mechanically expensive enough that few of the 86 studies we surveyed even attempted it.

AI facilitates visualization, but seeing remains a human act. AI can now perform the data transformations, filtering and formatting raw data, and even propose the data mappings that associate dimensions of that data with visual properties~\cite{card1999using}. What it cannot do is the rest of the pipeline: using human perceptual and cognitive skills to recognize the patterns that matter, and to judge whether they are genuine or artifacts of how the data was transformed. A visualization system should be built around that judgment, offering not one predefined view but the interaction that lets a developer move between them, from an overview down to the detail that confirms or overturns a pattern, and back again, one question guiding the next.
This reframes what a visualization system should be; a shift we develop further below.
The domains our extended survey found under-served were not a minor oversight; they were the ones where cheap data transformation and mapping had been missing, leaving the developer's perceptual and cognitive work as visualization's one irreducible contribution.
Therefore, \emph{we should treat previously under-served domains (i.e., rationale, contracts, and policies) as priority targets for new design studies now that AI removes their main barrier: the cost of preparing the data. 
The scarcity that once limited attention to these domains no longer holds}.


\subsection{The tool-lifespan problem AI may worsen before solving}
We reported that visualization tools from our sample had an average lifespan of 3.7 years, and that only 26\% made source code publicly available. We read this, at the time, as evidence of a field that produces prototypes rather than infrastructure, but the deeper cause is not technical; it is the incentive structure. Most of these tools were built by graduate students to validate a specific hypothesis for a specific publication, and once that goal is met, there is no comparable incentive to document, generalize, or maintain the result. Reproducibility is treated as an ethical obligation rather than a rewarded outcome, and prototypes are abandoned accordingly. Generative AI does not fix this incentive structure, and its non-deterministic nature threatens to make the underlying problem worse, not better. If producing a plausible-looking visualization tool becomes nearly free, the same incentive to abandon it after publication remains, while the volume of abandoned, undocumented, and unreproducible artifacts could grow by orders of magnitude, each one harder to audit precisely because its behavior was never fully deterministic or specified to begin with. What was a sustainability problem measured in dozens of short-lived research prototypes could scale to millions of ungoverned generated artifacts as generative tooling makes the debt exponentially larger.
Therefore, \emph{incentives for documentation and maintenance, not just novelty, must be built into how we evaluate and reward research contributions. A tool that is smaller in scope but well-documented and maintained should count as a stronger contribution than a more ambitious one that will not survive its authors' graduation}.


\subsection{Medium: a prediction that held}
To check our 2016 prediction against the decade that followed, we scanned the titles of every paper published at VISSOFT from 2017 through 2025, across all tracks, rather than restricting ourselves to design studies as our 2016 methodology had. Papers explicitly naming an immersive 3D environment, virtual reality, or augmented reality as their delivery medium grew from roughly one in a hundred to roughly one in seven; we found no comparable growth in explicit mentions of wall displays, 3D glasses, or tablets, though we did not track those categories as systematically. This is a title-level count, not a full re-read of each paper, so it almost certainly undercounts studies that use an immersive medium without mentioning it in the title; the comparison to other media should be read as suggestive, not definitive.

Part of that earlier scarcity was logistical: building CityVR took months on a young HTC Vive toolchain~\cite{merino2017cityvr}, and CityAR was harder still, with sparse HoloLens documentation consuming most of the project~\cite{merino2018overcoming}, the ordinary cost of academic AR/VR research without industrial resources.

That barrier has largely dissolved, and we read the roughly fifteen-fold increase in immersive-medium papers as evidence of exactly that. The same class of prototype that took us months in 2016, and that pushed us toward the more mainstream HTC Vive over HoloLens precisely to reduce this friction, could plausibly be built in a fraction of the time today, with AI-generated scaffolding, boilerplate, and device-specific idiosyncrasies absorbing most of the effort that once fell to the developer. Time previously spent fighting an unfamiliar SDK can now be directed towards more robust prototypes, richer interaction design, and the kind of documentation and maintenance that our field has chronically underfunded. Read together with that argument, the opening matters more than the convenience: the same AI-generated development that threatens to worsen the tool sustainability problem is, in the same breath, the most plausible way we currently have to address it.

If a new medium emerges next, it will likely not be another headset. Smart glasses are approaching the market not as vehicles for augmented or mixed reality, but as ambient, AI-mediated access to visual and auditory content, glanceable rather than immersive. For software visualization, that reframes the question entirely: not which headset renders a system best, but what it means for a developer to carry a running conversation with their systems everywhere, glancing at a dependency graph on a commute or a build failure mid-conversation, visualization woven into the day rather than opened as a destination. We cannot yet answer what that visualization should look like, only flag it as the medium shift this survey's next decade should be watching for.
Therefore, \emph{medium should become a first-class, explicitly reported dimension of every design study, not an afterthought; the field cannot track a shift it does not consistently measure}.


\section{Vision: Toward a Perceptual Research Agenda}
\label{sec:vision}
If matching developer needs to techniques is no longer the bottleneck, what should the field optimize for instead? We propose three shifts, each following directly from the reflections above.

\subsection{From technique catalogs to perceptual capacity}
The community's dominant unit of contribution has been the individual technique: a new layout, a new metaphor, a new tool, evaluated against a specific task. This made sense when producing a visualization was the expensive step. Today, AI can generate a competent visualization in response to a single question. We propose that VISSOFT research redirect part of its effort from designing more techniques toward strengthening the human side of the loop, focused on what makes software specifically hard to see. For instance, what makes the generated view of a dependency graph, a commit history, or an agent's reasoning trustworthy at a glance, and how quickly a developer can detect that such a view is misleading rather than merely unfamiliar. This is not an invitation to import perceptual psychology wholesale, that territory belongs to the broader visualization community; it is a call to ground perceptual research in the artifacts and questions software uniquely produces, the kind of grounding VISSOFT has always supplied and generic visualization research cannot.

\subsection{From monolithic tools to disposable, continuous seeing}
We described earlier how AI, by automating data transformation and mapping, can make a long-neglected domain like rationale tractable for the first time, leaving the developer to recognize and judge the patterns that matter. The natural end state of that shift is architectural, not just topical. Instead of a small number of task-specific tools, each built to answer one predefined question about one kind of artifact, we expect systems organized around a general capability: continuously examining a system from whatever angle the current question demands. Each view answers the question in front of it, then is abandoned the moment a new one arises, the same shift from artifact to channel, now pushed to its architectural conclusion and already practiced, not merely proposed, in AI-assisted moldable development environments such as Glamorous Toolkit~\cite{nierstrasz2024moldable}.

This does not mean a general-purpose model should do all of the work. Transforming data and mapping it to visual properties is a solved, deterministic problem; reinventing it with a language model on every request wastes compute and energy for no gain in quality. Generative AI earns its place elsewhere, and more efficiently: translating a request that starts out as an intuition, not yet articulated, into a specification deterministic machinery can execute reliably only if that machinery already knows which representation fits which kind of question.

That is exactly the knowledge the VISSOFT community has accumulated over decades, and it is what a general-purpose model lacks on its own: not the ability to produce code for a chart, but the grounding to know which chart the question calls for. Turning that accumulated knowledge into the training and grounding substrate for on-demand visualization is the field's opening, not a threat to its relevance.

\subsection{A new object of study: visualizing the reasoning of AI itself}
Visualization for AI is already an active research area, with substantial progress in explaining model internals, attention mechanisms, and decision boundaries~\cite{hohman2018visual}. That progress targets the model as the object of understanding, not the software it produces. Classical software visualization targets the other side of that gap: code that grew illegible over years of undocumented human change, but that was, at some point, written by someone who understood it. AI-generated code fits neither category. It was never authored with human comprehension as a goal at all; it emerges from a statistical process trained on human examples, which makes it often readable, but not reliably so, comprehensibility here is an emergent side effect of training, not a guarantee left by an author, and it can vary from one generated fragment to the next. We propose that AI-generated code, and the process that generates it, become a first-class object of study for software visualization specifically, one that reveals not only what the code does, but how consistently it can be trusted to mean what it appears to mean.

The distinction that matters here is between AI-assisted coding, where a developer remains the author, reading, editing, and integrating each suggestion, and vibe coding, where the developer specifies intent and accepts output with little line-by-line review. An AI with no reader in mind has no incentive to write legible code. The result can be functionally correct and structurally opaque at once: a system nobody, human or otherwise, wrote with comprehension in mind. This is a genuinely new problem for our field, distinct from the reverse-engineering legacy-code problem that motivated much of the software visualization literature we surveyed in 2016, because the artifact was never legible to begin with, rather than having become illegible through years of undocumented change.

We see two concrete openings. First, software visualization techniques developed for legacy comprehension, dependency graphs, call graphs, and execution traces are directly applicable to vibe-coded systems and may, in fact, be more urgently needed there since the usual scaffolding of comments, meaningful naming, and incremental human familiarity is often absent. 
Second, visualization can move from a post-hoc comprehension aid to a component embedded inside the vibe-coding loop itself: reifying an agent's tool calls, intermediate steps, and proposed diffs as small, navigable tools lets a developer inspect what the agent is doing before accepting a change, turning acceptance from a leap of faith into an informed judgment. Moldable development is one effort already moving in this direction~\cite{nierstrasz2022making}.\footnote{Demonstrated in the Moldable Chat integration of LLMs into Glamorous Toolkit: \url{https://youtu.be/y0FQrVq74BI} (accessed July 2026).}
In both cases, an agent's reasoning, its discarded alternatives, its intermediate hypotheses, and the tool calls it made along the way are themselves a data source with no obvious analog in our original 2016 taxonomy. It is a data source our field is well positioned to take on, precisely because reading structure out of complex, undocumented artifacts is what software visualization has always done.





\section{Risks and a Closing Reflection}
\label{sec:risks}

The growing use of AI-generated visualizations on the fly introduces the same reliability concerns as other AI-generated outputs. A generated chart can misrepresent its underlying data just as a generated sentence can misdescribe its source. What gives a visualization its evidentiary edge is not new: Shneiderman's overview-first, details-on-demand mantra~\cite{shneiderman1996eyes} reminds us that a view is only as trustworthy as its connection to the data it summarizes. What is new is that generation is now automated and cheap enough that maintaining this connection can no longer be assumed; it has to be designed in deliberately. A generated view should be built from simple, directly checkable mappings to measurable values and remain navigable down to the source data, not merely become more numerous.

This is not a reason to retreat from disposable, on-demand visualization. It is a reason to enforce explicitly the principles the field has known for decades. The value of navigability extends beyond human perception: the same structured, drillable access that lets a developer trace a view back to its source data can also let an agent query and traverse that structure before presenting a change, checking its own work against the evidence a human would consult.\footnote{Based on early work on integrating LLMs into Glamorous Toolkit; Oscar Nierstrasz, personal communication, 2026.}

This shift also calls for benchmarks that assess not whether a system can produce a visualization, but whether the resulting view remains actionable: faithful to its underlying data, traceable to its sources, and supportive of reliable reasoning.

A second, more disciplinary risk is competing with machines on terrain where they already win. The durable response is to build around capacities with no machine analog: perception, ethical judgment, and empathy, rather than treating them as noise to engineer around.

Both risks share a common shape: AI relocates verification and judgment rather than eliminating them. An unverified claim or ungrounded chart can now enter a codebase with far less friction than a decade ago. This is why our agenda centers on strengthening human perceptual and evaluative capacities rather than producing more techniques: generation can be delegated; checking what is shown against what is true cannot.

We titled our 2016 survey \emph{Towards} Actionable Visualization, a title that was, at the time, as much an admission as an ambition: the field had techniques, but developers still lacked an easy way to reach for them, leaving actionable visualization mostly aspirational.
The barrier we described then was not conceptual; we knew what a useful visualization looked like. It was the cost of building one for a specific, momentary question, a cost too high to justify anything short of a sustained research effort. This paper has argued that this barrier is now falling. 
This reduction in cost allows a visualization to be generated, inspected, and discarded in service of a single question, rather than requiring its own multi-year research agenda.
If the visualization a developer needs can be produced as cheaply as the question that motivates it, then reaching for one stops being a decision that competes with the task at hand and becomes part of it. \emph{Towards} was the right word for 2016. We think the field is close enough now to drop it, and we recognize the standard that follows: not more research toward actionable visualization, but visualization that is, finally, actionable.

\bibliographystyle{IEEEtran}
\bibliography{vision}
\end{document}